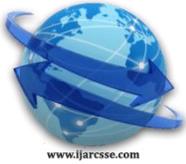

# UML modelling of geographic routing protocol "Greedy Perimeter Stateless Routing" for its integration into the " Java Network Simulator''

| Mohammed ERRITALI | Oussama Mohamed Reda | Bouabid El Ouahidi |
|---|---|---|
| *Department of Computer Science, Faculty of Sciences, Mohamed V University Rabat Morocco* | *Department of Computer Science, Faculty of Sciences Mohamed V University Rabat Morocco* | *Department of Computer Science, Faculty of Sciences Mohamed V University Rabat Morocco* |

*Abstract*— In this work we propose an UML modeling of the "Greedy Perimeter Stateless Routing" (GPSR) protocol that integrate this geographic routing protocol, into "JavaNetwork Simulator" to simulate and study this protocol in a first time and offer some improvement in these features. Java Network Simulator (JNS) is a project of "translation" of Network Simulator (NS) in Java initiated by "the UCL Department of Computer Science." This simulator is not as complete as ns-2, but it is much more accessible to programmers unfamiliar with Tcl. Java Network Simulator does not support so far, no routing protocol for vehicular ad hoc networks and all the routing decisions are made statically or using RIP and OSPF. By modeling and integrating the routing protocol GPSR to JNS, users will be able to understand the concept of the geographic routing and how the routing information is transmitted and updated between nodes in vehicular ad hoc network. The article first examines the architecture of the Java Network Simulator, then gives a brief review of the routing protocol GPSR and finally presents our UML modeling incorporating GPSR in the Java Network Simulator.

## I. INTRODUCTION

Vehicular ad hoc networks (VANET) are networks whose topology does not receive any pre-existing infrastructure. It is formed according to the appearance and movement of vehicles.In these networks the routing problem is to find a path from a data source to its final destination, through a series of intermediate nodes in spite of the rapid change in network topology caused by movement of nodes (vehicles). These networks require a reactive routing algorithm that finds valid routes, faster than the topology changes, which do not lead the network in a state of congestion when they learn of new routes.

The geographic routing seems to be an ideal candidate for VANET networks for several reasons. First, this type of routing allows scalability. In addition, it benefits from the availability of inexpensive GPS receivers on the market. Thus, many geographic routing protocols have been proposed in recent years like GPSR "Greedy Perimeter Stateless Routing" [1,3,5].

In this context to study the dynamic behavior of vehicular ad hoc networks lead us to a set of experiments. However, this solution is inapplicable in practice because of the construction costs of such an experimental network.

An alternative is to introduce the simulation, which allows tracing the behavior of the network. Thus, it allows us to understand the network behavior and explore scenarios with a variety of configurations.

Java Network Simulator (JNS) [4] is a project of "translation" of Network Simulator (NS) in Java initiated by "the UCL Department of Computer Science." This simulator is not as complete as ns-2, but it is much more accessible to programmers unfamiliar with Tcl.

Java Network Simulator only supports up to now no routing protocol for ad hoc networks and all the routing decisions are made statically or using RIP and OSPF.

The aim of the project "JNS" is to develop an interactive simulation system for teaching purposes. JNS supports the data link layer, network and transport layers and allows the simulation of protocols above the IP layer.

Our objective in this work is to integrate a geographic routing protocol, Greedy Perimeter Stateless Routing in "Java Network Simulator" to simulate and study this protocol in a first time to offer some improvement in these features.

By integrating the routing protocol GPSR to JNS, users will be able to understand the concept of geographic routing and how the routing information is transmitted and updated between the nodes in the vehicular ad hoc network. In addition, the implementation of the GPSR protocol will improve the functionality of the simulation environment.

The first part of our work deals with the architecture of JNS and its way of routing packets. The second part summarizes the geographic routing protocol GPSR. Finally we suggest a UML GPSR protocol allowing integration at the JNS.

## II. ARCHITECTURE OF JNS

Java Network Simulator project is a "translation" of Berkeley Network Simulator (ns) in Java. In JNS the main class is the class Simulator, which contains the queue event simulator, and the run method that allows initiated a simulation experiment. In addition to this there are five packages: element, command, agent, trace and util [2].

1. **jns.element**: contains the static elements of a network (eg, nodes, links, interface, etc.).

2. **jns.agent**: contains interfaces for agents. A new protocol to JNS must implement one of three interfaces: Agent, CL_Agent or CO_Agent. CO_Agent and CL_Agent inherit from agent. CL means a connection-oriented service and CO means a not connection-oriented service.

3. **jns.trace**: provides the functionality for tracing what is happening in the network. A class converts all events in a trace file, which is then introduced into Javis. Users can then view the network and its behavior.

4. **jns.command**: contains classes used to schedule calls to functions of the simulator. JNS will call a function in this class, when a command should be executed.

5. **Jns.util**: contains utility classes. These include the queue of data, a priority queue, the class ipaddr, RoutingTable class, the Preferences class...

These classes are used by some classes in the package jns.element, for example, the class IPHandler [4].
The routing protocol GPSR operates at the network layer, it is necessary to understand how IP packets are transferred by JNS for the implementation of the GPSR protocol and its integration into JNS. Thus, we will illustrate the mechanism for transferring data in JNS.

The following classes in the package jns.element are involved in the transfer of IP packets:
• IPHandler, the generator and receiver of IP packets.
• SimplexInterface, place the incoming and outgoing packets in the queue and sends either a Link or IPHandler.
• The SimplexLink, which holds the packets for a certain period of time (delay) before transmitting them to other interfaces.

Before a packet can be sent to the network layer, the above elements must be linked first. JNS provides a function called attach () to link the lower level agents with the higher level agents. When this function is called, the lower level agent, SimplexInterface for example, will be fixed inversely to the higher level agent in IPHandler for example. This function allows the agent higher level to obtain a reference to the lower level agent.

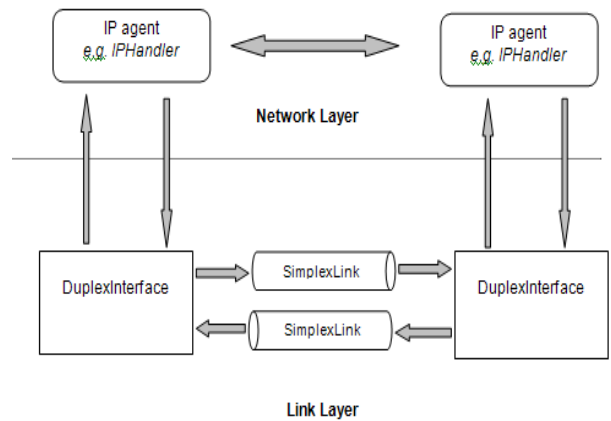

Fig 1: JNS architecture

## III. GREEDY PERIMETER STATELESS ROUTING

In vehicular ad hoc networks (Vehicular Ad Hoc Networks), the routing problem is to find a path from a data source to its final destination, through a series of intermediate nodes in spite of the rapid change in network topology caused by movement of nodes (vehicles).

These networks require a reactive routing algorithm that finds valid routes quickly as topology changes. However, the limited capacity of the channel network requires routing algorithms, which do not lead the network in a state of congestion when they learn of new routes.

For this reason, most routing protocols designed for such networks are geographic routing protocols because they can prevent overload of information exchanged between the nodes that seek to obtain the network topology and routing tables. These geographic routing protocols are based on the fact that all nodes know their position, for example, with GPS equipment (Global Positioning System) or by a positioning system distributed.

Greedy Perimeter Stateless Routing, GPSR is a reactive and efficient routing protocol for vehicular ad hoc networks. In contrast to routing algorithms implemented before, using the concepts of graph theory, the shortest path and transitive accessibility to find routes, GPSR exploits the correspondence between the location and connectivity in a wireless network, using the positions of the nodes to make packet forwarding decisions.

In a VANET network, nodes are liable to move. It is therefore necessary to use a mechanism that allows each node to know the position of its neighbors. To signal their presence and location nodes flood the network by sending a packet signaling (message "beacon") that contains the position and an identifier. The periodic exchange of these packets allows nodes to construct their position table. The period for issuing messages "beacon" depends on the rate of mobility in the network and the radio range of nodes. Indeed, when a node does not receive a message "beacon" of a neighbor after a time T, it considers that the neighbor in question is no longer in its coverage area and removes it from his position table . One advantage of message "beacon" is that each node only needs information about its direct neighbors, which requires little memory. Alternatively, the GPSR protocol allows the node to encapsulate a few bits of their position in the data packets it sends, "We encode position as Two Four-byte floating point quantities, for x and y coordinate values."[3]. In this case, all interfaces of the nodes must be in promiscuous mode to receive packets if they are in the coverage area of the issuer.

The routing of GPSR packets is done in two modes according to the density of the network: the "Greedy Forwarding" and "Perimeter Forwarding" (respectively called GF and PF in the following).

*3.1 Greedy Forwarding*

The GF constructs a road browsing the nodes from the source to the destination where each node receives a packet that forwards it by a jump to the intermediate node closest to the destination in its coverage area. Figure 2 shows an example of this mode of transport [3].

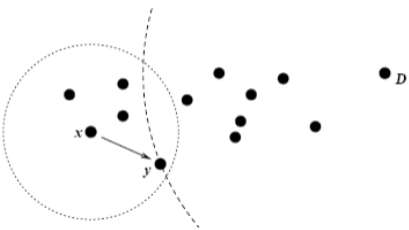

Fig2: y is the neighbor of x closest to the destination D.

The method "Perimeter Forwarding" is used when a node does not find any neighbor closer than him to the destination or destination is not within reach of it. Figure 3 shows an example of this routing [3].

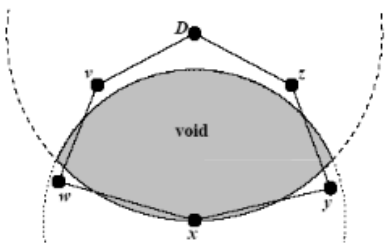

Fig 3: X is closer to that of its neighbors y, w.

*3.2 Perimeter Forwarding*

The method "Perimeter Forwarding" consists to transform the network topology in a planar graph (not containing of edges that intersect). This graph can be either (Relative Neighborhood Graph) RNG or (Gabriel Graph) GG (Figure 4).

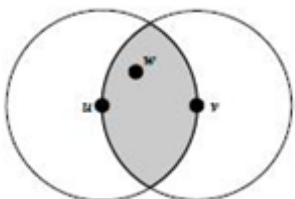

Figure 4.1: Relative Neighborhood Graph

RNG: the node u considers that v belongs to the graph RNG if the hatched area is empty.

$$\forall w \neq u,v: d(u,v) \leq \max[d(u,w), d(v,w)]$$

In this case (x, v) is an edge of the graph.

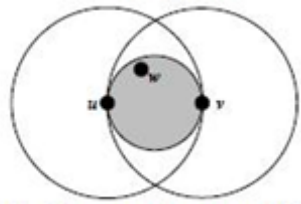

Figure 4.2: Gabriel Graph

GG: the node u considers that v belongs to the graph GG if the hatched area is empty.

$$\forall w \neq u,v: d^2(u,v) < [d^2(u,w) + d^2(v,w)]$$

In this case (x, v) is an edge of the graph.

Then the packet traverses the graph to the destination using the right hand rule defined as follows: When a packet arrives at a node x from node y, the way forward is the next node that is in the opposite direction of clockwise starting from x and the segment [xy] while avoiding the "crossing links" (road already traveled).

Now we describe the GPSR protocol by combining the two routing methods: A GPSR packet contains a header field for the routing mode. This field contains "Greedy» when the routing is "greedy forwarding" and "Perimeter" when routing is "Perimeter forwarding". A node x receives a packet in "Greedy" mode examines the table of neighbors. If it finds the nearest neighbor of the destination, it transmits the packet to this neighbor. Otherwise, the node will change the mode field of the header of the packet with "Perimeter" and record its location. Then it constructs a planar graph of its neighbors and transmits its packet through this graph. Figure 5 shows an example of this mode of transport.

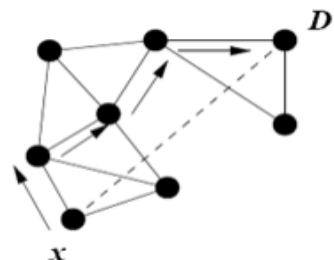

Fig 5: Perimeter forwarding. D is the destination; x is the node where the packet enters in Perimeter mode.

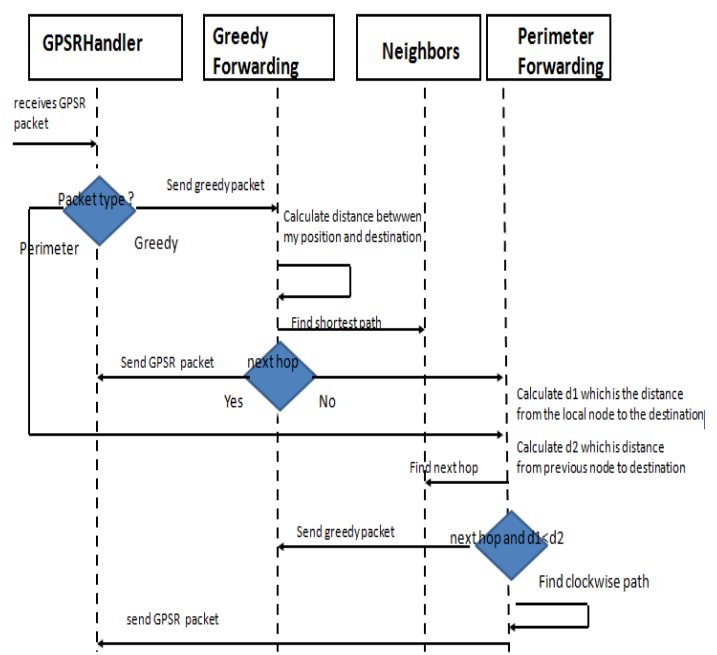

## IV. DESIGN OF GPSR PROTOCOL

### 4.1 Design

The classes that we identified to implement the prototype of GPSR are GPSRHandler, Neighbors, Beaconing, Greedy Forwarding and Perimeter Forwarding. These classes will be introduced in the following sections.

#### 4.1.1 The GPSRHandler class:

The generator and receiver of GPSR packets.

#### 4.1.2 The Neighbors class:

The role of this class is to maintain the location information of all the immediate neighbors.
It also allows to add, delete and update the status of neighbors, and finally to build the topology GG or RNG.

#### 4.1.3 The Beaconing class

Beaconing class is used to send beacon request packet and response, send periodic beacon packet to neighbor nodes and check regularly the connections to neighboring nodes.

#### 4.1.4 The Greedy Forwarding class

This class allows you to send and receive packets Greedy Forwarding mode and find the nearest way over to this mode of transport.

#### 4.1.5 The Perimeter Forwarding class

This class allows you to send and receive packets Perimeter Forwarding mode and find the path that follows the rule of law "Right-Hand Rule."

### 4.2 sequence diagrams

#### 4.2.1 Beacon algorithm

The algorithm allows a beacon node to have the locations of its neighbors. Periodically, each node sends a beacon containing its own identifier and location by using two four-byte floating point values for x and y. If a node doesn't receives a beacon packet from a neighboring node after a certain period of time, the GPSR router assumes the neighbor is gone and will remove it from the table of valid neighbors.
The sequence diagram below illustrates the steps of the beacon algorithm:

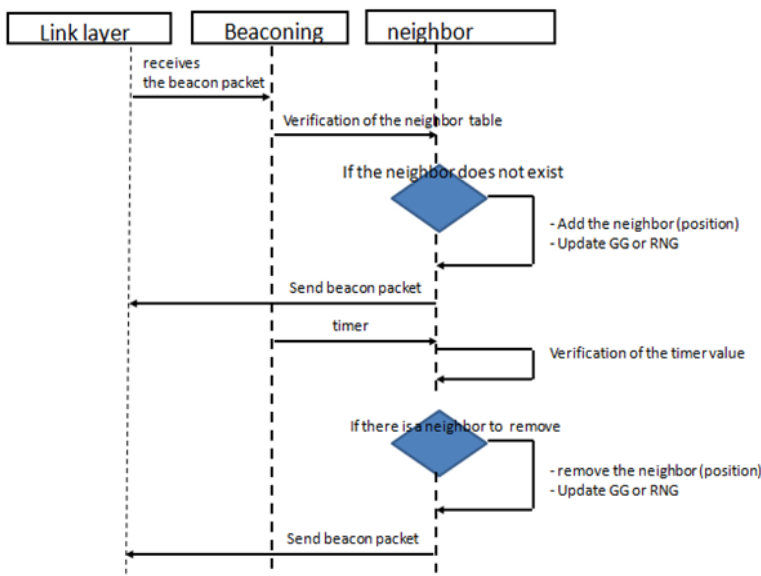

#### 4.2.2 GPSR Data packet transfer

GPSR allows nodes to know which are its closest neighbors to the destination (using beacons). To calculate a trajectory, GPSR uses greedy forwarding algorithm that will send information to the final destination using the most efficient path possible. If the transfer fails using greedy forwarding, the perimeter forwarding that will be used.
The sequence diagram below illustrates the steps of GPSR Data packet transfer:

#### 4.2.3 Intgration of GPSR protocol in JNS

In this section we will illustrate the mechanism of data transfer using the GPSR in JNS is using the diagram presented in the following figure:

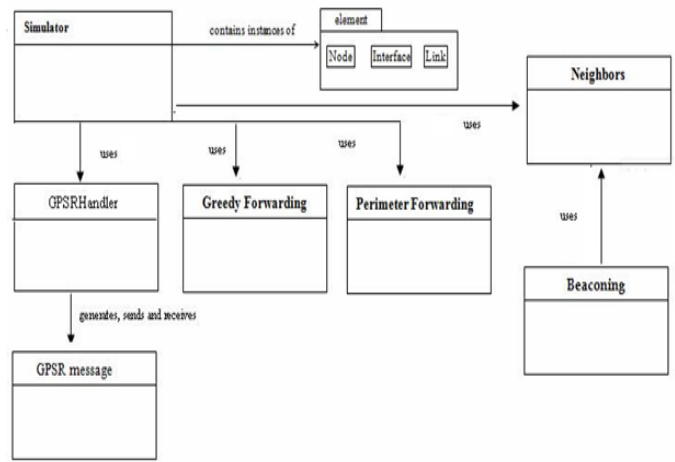

## V. CONCLUSIONS

The problems of communication and safety in vehicular ad hoc networks attract more and more attention from research groups.
Indeed, the ease of deployment of vehicular ad hoc networks and their spontaneous nature makes them a compelling solution for security and comfort of motorists and their passengers.
However, the VANET have basic properties that make it difficult to deliver the correct data from a source node to a destination node: These networks require that all users work together to route information of other users on the same network VANET. This hypothesis and several other characteristics (mobility, bandwidth ...). Make the problems of routing and security of communications in these networks a capital axis of research.
In this paper we presented a UML modeling protocol GPSR and architecture of JNS.
As prospects of this work, we will use our UML to implement and test the geographic routing protocol, Greedy Perimeter Stateless Routing for Vehicular Ad Hoc Network, in the "simulator Java Network Simulator" in a first time before proposing some improvement to these features, then go take a integration of the standard IEEE802.11 p which has the physical layer and MAC for a complete simulation of VANET network within the JNS.

## REFERENCES

[1] *http://www.icir.org/bkarp/gpsr/gpsr.html*
[2] *Kelvin Jian-Ling PENG, Implementation of the OSPF protocol with the Java Network Simulator, MSc project, University College London, 1999*
[3] *Karp, B. and Kung, H.T., Greedy Perimeter Stateless Routing for Wireless Networks, in proceedings of the Sixth Annual ACM/IEEE International Conference on Mobile Computing and Networking (MobiCom 2000), Boston, MA, August, 2000.*
[4] *ftp://cs.ucl.ac.uk/nets/src/jns/jns-1.6/doc*
[5] *http://www.memoireonline.com/04/10/3394/m_Greedy-perimeter-stateless-routing-sur-omnet2.html*